%% file: chic.tex
\documentclass[12pt]{article}
\usepackage{graphicx}
\usepackage{color}
\usepackage{multirow}

\newcommand\pubdate{\today}

\def\Journal#1#2#3#4{{#1} {\bf #2}, #3 (#4)}

\def\usc{Department of Physics and Astronomy\\
  University of South Carolina, Columbia, SC 29208, USA}
\def\support{\footnote{Speaker on behalf of Belle Collaboration, supported by U.S. Department of Energy.}}
\def\PLB{{ Phys. Lett.}  B}
\def\PRL{ Phys. Rev. Lett.}
\def\PRD{{ Phys. Rev.} D}

\def\chicx{\chi_{cJ}}
\def\NIMA{{ Nucl. Instrum. Methods Phys. Res., Sect. A }}

\def\pccm{p^*_{\chi_{c\rm{J}}}}

\def\Title#1{\begin{center} {\Large #1 } \end{center}}
\def\Author#1{\begin{center}{ \sc #1} \end{center}}
\def\Address#1{\begin{center}{ \it #1} \end{center}}

\newenvironment{Abstract}{\begin{quotation}  }{\end{quotation}}
\newenvironment{Presented}{\begin{quotation} \begin{center} 
             PRESENTED AT\end{center}\bigskip 
      \begin{center}\begin{large}}{\end{large}\end{center} \end{quotation}}
\def\Acknowledgements{\bigskip  \bigskip \begin{center} \begin{large}
             \bf ACKNOWLEDGEMENTS \end{large}\end{center}}

\input econfmacros.tex

\begin{document}
\begin{titlepage}
\hfill \pubdate

\vfill
\Title{Study of $B \to \chi_{cJ} X$ at Belle}
\vfill
\Author{V. Bhardwaj\support}
\Address{\usc}
\vfill
\begin{Abstract}
In spite of the fact that the two-body $B$ decays into $\chi_{c2}$ such as $B \to \chi_{c2} K^{(*)}$ are suppressed by the QCD factorization effect, the inclusive $B \to \chi_{c2} X$ branching fraction amounts to one third of the non-suppressed $B \to \chi_{c1} X$ decays because of the decay modes to the multi-body final states. Using a large statistics $\Upsilon(4S)$ data sample corresponding to 772 million $B$ meson pairs accumulated by the Belle detector at the KEKB $e^+e^-$ collider, precise measurements of inclusive $B \to \chi_{c1}$ and $\chi_{c2}$ branching fractions are carried out. The multi-body final states such as $\chi_{cJ} K \pi$, $\chi_{cJ} K \pi \pi$ and so on are also investigated to look for new charmonium-like resonance. 

\end{Abstract}
\vfill
\begin{Presented}
The 7th International Workshop on Charm Physics (CHARM 2015)\\
Detroit, MI, 18-22 May, 2015
\end{Presented}
\vfill
\end{titlepage}
\def\thefootnote{\fnsymbol{footnote}}
\setcounter{footnote}{0}
%
\section{Introduction}

Inclusive production of $\chi_{c2}$ mesons in $B$ decays is relatively 
large~\cite{Belle_PRL_89_011803_2002, BaBar_PRD_67_032002_2003}
in spite of the fact that two-body $B$ decays into $\chi_{c2}$ are
highly suppressed~\cite{Belle_Soni_2006,BaBar_PRL_102_132001_2009,Belle_PRL_107_091803_2011}
(due to the angular momentum conservation).
Differential branching fraction ($D\mathcal{B}$) in
bins of $\chicx$ ($J=1,2$)~\cite{CX} suggests that $\chi_{c2}$ is found to 
be coming from three-body or higher multiplicity 
decays~\cite{Belle_PRL_89_011803_2002, BaBar_PRD_67_032002_2003}, 
which have not been studied in detailed yet. More experimental input is 
needed to study these multi-body decay modes. 

Study of more than three-body $B$ decay modes with $\chi_{c1}$ and $\chi_{c2}$ 
in final state not only help in understanding $B$ meson decays but also
provide portal to search for charmonium/charmonium-like exotic states in
one of the intermediate final states.
For example, looking at the $\chi_{c1} \pi^+\pi^-$ invariant mass
spectrum in $B \to \chi_{c1} \pi^+\pi^- K$ decays,  one can search for
$\chi_{c1}(2P)$ and/or $X(3872)$.

Using the $\chicx \to J/\psi \gamma$ modes, we report on the 
inclusive branching fraction ($\mathcal{B}$) of $B \to \chicx X$ 
decays. To further understand $\chi_{c1}$ and $\chi_{c2}$ production 
in $B$ decays, 
we reconstruct the following exclusive $B$ decays: 
$B^0 \to \chicx \pi^- K^+$, 
$B^+ \to \chicx \pi^+ K_S^0$, 
$B^+ \to \chicx \pi^0 K^+$, 
$B^+ \to \chicx \pi^+ \pi^- K^+$ and
$B^0 \to \chicx \pi^+ \pi^- K_S^0$~\cite{mixchg}. 

\section{Data sample and event selection}
We use a data sample of  $772\times 10^{6}$ $B\overline{B}$  events 
collected with the Belle detector~\cite{abashian} at the KEKB asymmetric-energy 
$e^+e^-$ collider~\cite{kurokawa} operating at the $\Upsilon(4S)$ resonance.
All results presented here are preliminary. 


The $J/\psi$ meson is reconstructed via its decays to $\ell^+\ell^-$ 
($\ell =$ $e$ or $\mu$).  To reduce the radiative tail in the $e^+e^-$ mode, 
the four-momenta of all photons within 50 mrad with respect to the original
direction of the $e^+$ or $e^-$ tracks are included in the invariant mass 
calculation, hereinafter denoted as  $M_{e^+e^- (\gamma)}$. 
The reconstructed invariant mass of the $J/\psi$ candidates is required 
to satisfy 2.95 GeV$/c^2 < M_{e^+ e^-(\gamma)} < 3.13$ GeV$/c^2$ or 
3.03 GeV$/c^2 < M_{\mu^+ \mu^-} < 3.13$ GeV$/c^2$. 
For the selected $J/\psi$ candidates,  a vertex-constrained fit is
applied and then a  mass-constrained fit is performed in order to improve 
the momentum resolution.   

The $\chi_{c1}$ and $\chi_{c2}$ candidates are reconstructed by combining
 $J/\psi$ candidates with a photon having energy ($E_{\gamma}$)  
larger than 100 MeV in the laboratory frame.  To reduce the combinatorial
background coming from $\pi^0 \to \gamma\gamma$, we use a likelihood function
that distinguishes an isolated photon from $\pi^0$ decays using the photon pair
invariant mass, photon laboratory energy and polar angle~\cite{koppenburg}.
We reject both $\gamma$'s in the pair if the $\pi^0$ likelihood probability 
is larger than 0.3 (0.8) for an inclusive study ($B \to \chicx K \pi$, 
$B \to \chicx K \pi \pi$ decay modes).

\section{Inclusive $B$ decays to $\chicx$}

\par
\begin{figure}[h!]
  \centering
  \includegraphics[height=60mm,width=80mm]{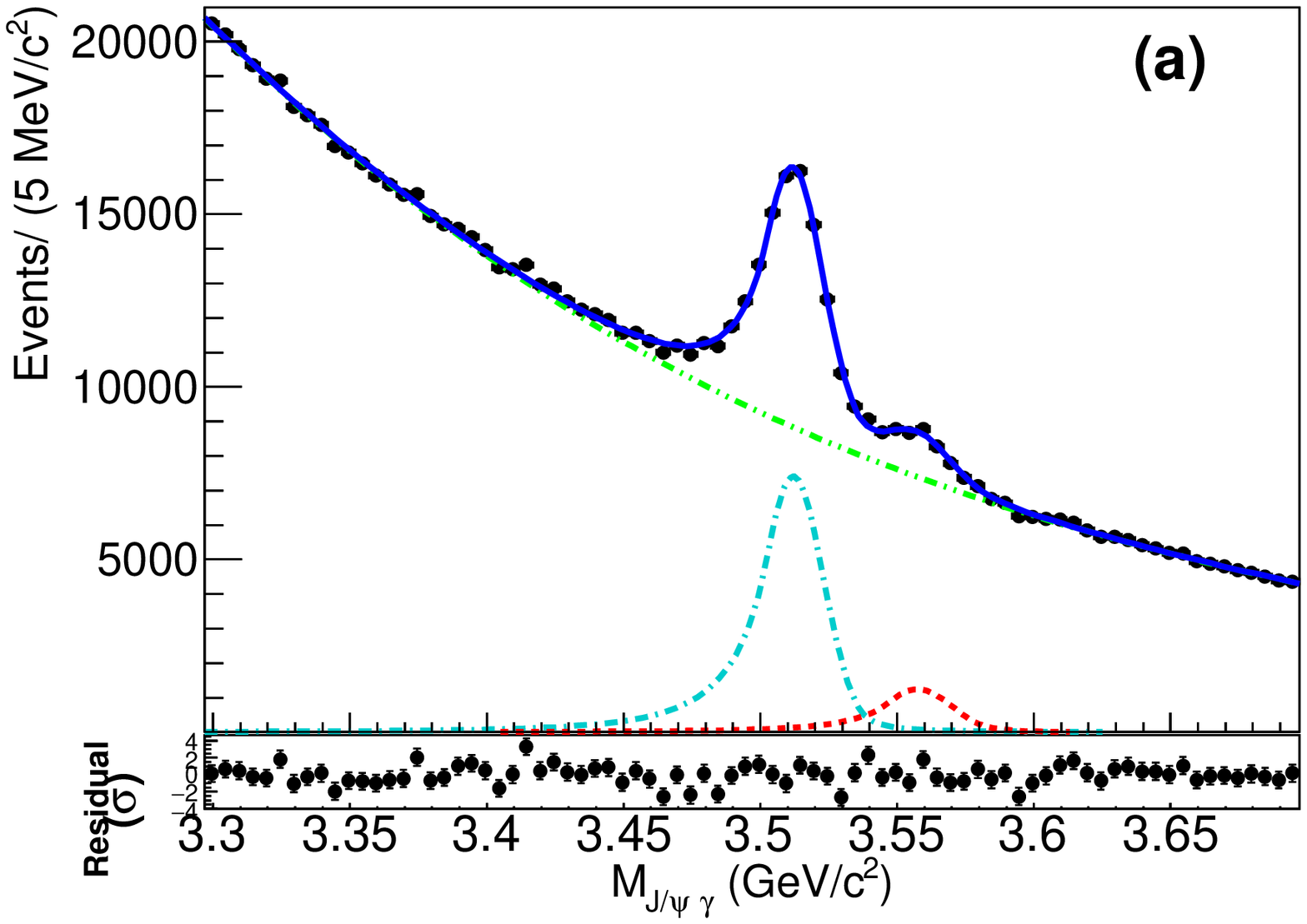}
  \includegraphics[trim=0.5cm 0.6cm 0.0cm 0.0cm,height=55mm,width=70mm]{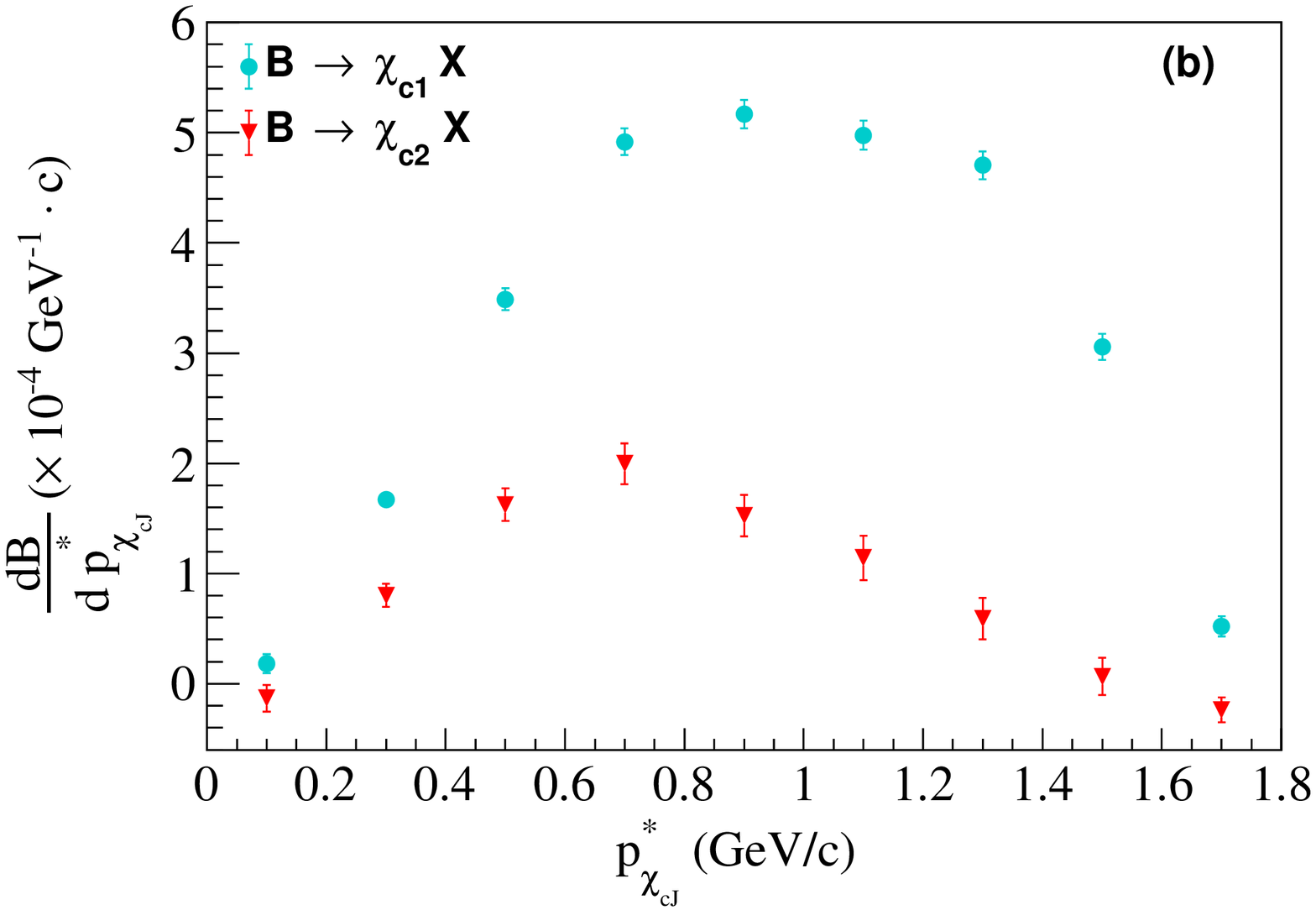}
  \caption{{(a)  $M_{J/\psi \gamma}$ distribution of the 
      $B \to \chicx(\to J/\psi(\to \ell^+  \ell^-)\gamma) X$ decays in 
      data. The curves show the signal 
      (cyan dash-dotted for $\chi_{c1}$ and red dashed for $\chi_{c2}$)
      and the background component (green dash-double-dotted for 
      combinatorial) as well as the overall fit (blue solid). The lower
      plot shows the pull of the residuals with respect to the fit.
      (b) Plots showing  $D\mathcal{B}(B \to \chi_{c1} X)$ and 
      $D\mathcal{B}(B\to\chi_{c2}X)$ in each bin of $\pccm$ for
      $B \to \chicx(\to J/\psi\gamma) X$.
      These plots are without feed-down subtraction.  The uncertainty shown in 
      these plots are statistical only. }}

  \label{fig:incdata_fit}
\end{figure}

 To identify the signal,  we use the $J/\psi \gamma$ invariant mass
$M_{J/\psi\gamma}$ distribution and extract the signal yield from a binned  
maximum likelihood fit. A double-sided Crystal Ball function is used to model 
the signal shapes of $B \to \chi_{c1} X$ and $B \to \chi_{c2} X$.
Figure~\ref{fig:incdata_fit} (a)  shows the fit to the $M_{J/\psi\gamma}$ 
distribution for $B \to \chi_{c1} X$ and $B \to \chi_{c2} X$ decays.


The reconstruction  efficiencies for the inclusive 
$B \to \chi_{c1} X$ and
$B \to \chi_{c2} X$ decays  are estimated to be 24.2\% and 25.9\%, respectively.
Uncertainty on the efficiency is estimated to be 4.0\%.

After subtracting the $\psi'$ feed-down contribution, we get the direct 
branching fractions 
 $\mathcal{B}(B\to \chi_{c1} X)$ and
 $\mathcal{B}(B\to \chi_{c2}X)$  to be
$(3.03      \pm 0.05 \pm 0.25)\times 10^{-3}$ and
$(0.70\pm 0.06\pm0.10)\times 10^{-3}$, respectively.
First (second) error is statistical (systematic).
Here, the systematic uncertainty dominates the measured branching fractions.

Figure~\ref{fig:incdata_fit} (b) shows the obtained distribution of
the  $D\mathcal{B}$ in bins of $\pccm$.
Suppression of two-body decay
of $\chi_{c2}$ is visible in the $\pccm$ distribution. Most of the
$\chi_{c2}$  production comes from decays of three-body or more.

\section{Exclusive reconstruction}
The reconstructed invariant mass of the $\chi_{c1}$  ($\chi_{c2}$) is 
required to satisfy 
3.467 GeV$/c^2 < M_{J/\psi\gamma} <$ 3.535 GeV$/c^2$ 
(3.535 GeV$/c^2 < M_{J/\psi \gamma} <$ 3.611 GeV$/c^2$). 
A mass-constrained fit is applied to the selected  
$\chi_{c1}$ and $\chi_{c2}$ candidates.

\begin{figure}[ht!]
  \includegraphics[trim=0cm 1cm 0cm 0.25cm,height=70mm,width=150mm]{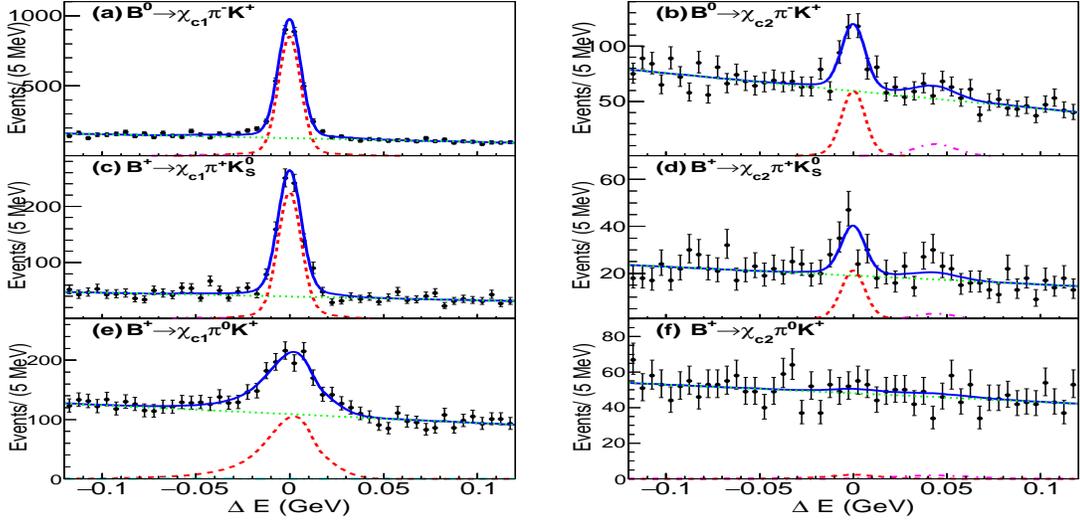}
  \caption{$\Delta E$ distribution 
    for (a) $B^0 \to \chi_{c1} \pi^- K^+$,
    (b) $B^0 \to \chi_{c2} \pi^- K^+$,
    (c) $B^+ \to \chi_{c1} \pi^+ K_S^0$,
    (d) $B^+ \to \chi_{c2} \pi^+ K_S^0$,
    (e) $B^+ \to \chi_{c1} \pi^0 K^+$, and
    (f) $B^+ \to \chi_{c2} \pi^0 K^+$ decay modes.
    The curves show the signal 
    (red dashed),
    peaking background (magenta dash-dotted) 
    and the background component (green dotted for 
    combinatorial) as well as the overall fit (blue solid).}
    \label{fig:B_chicJ_Kpi}
\end{figure}

\begin{figure*}
  \includegraphics[trim=0cm 1cm 0cm 0.25cm,height=70mm,width=150mm]{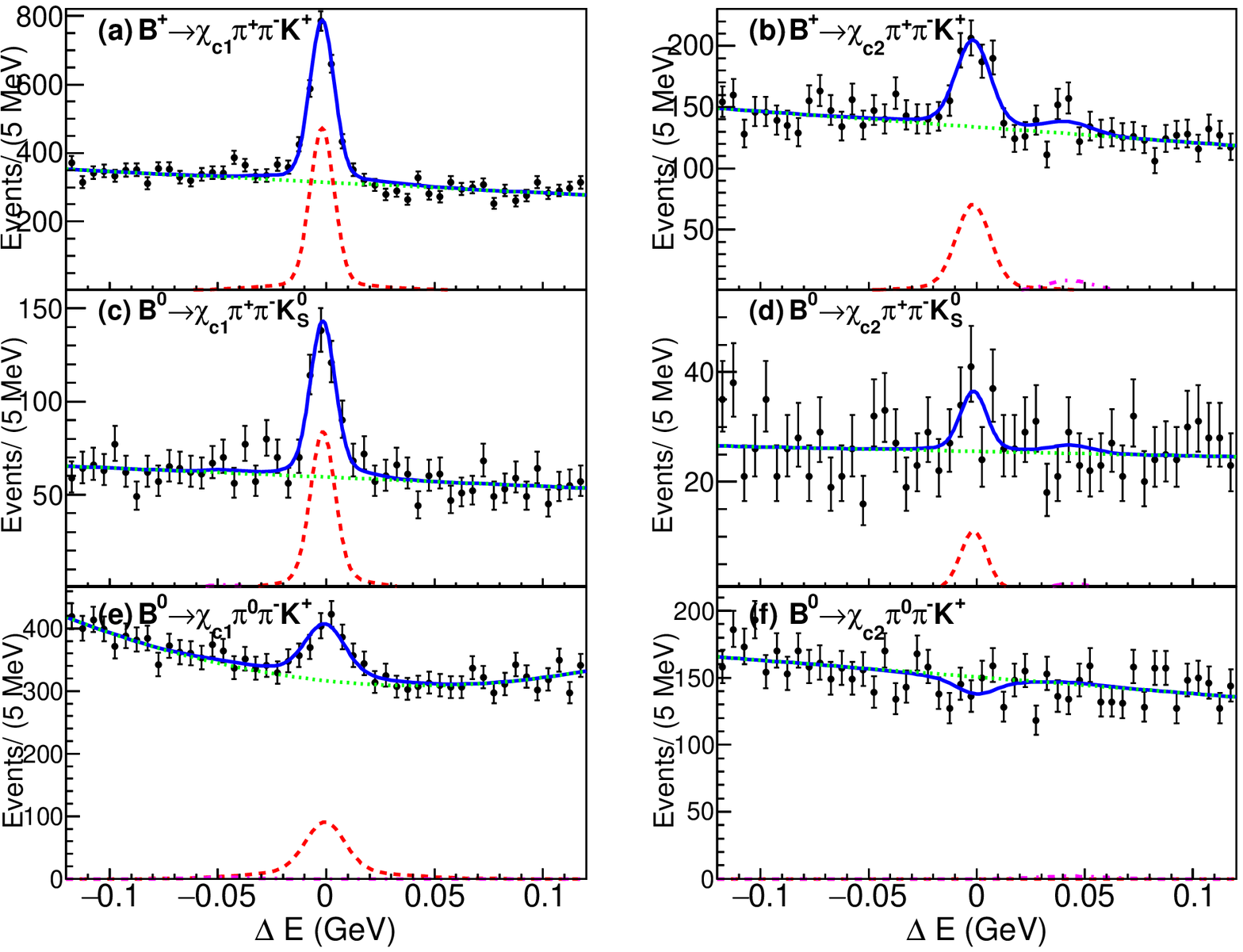}
  \caption{$\Delta E$ distributions 
    for (a) $B^+ \to \chi_{c1} \pi^+ \pi^- K^+$,
    (b) $B^+ \to \chi_{c2} \pi^+ \pi^- K^+$,
    (c) $B^0 \to \chi_{c1} \pi^+ \pi^- K_S^0$,
    (d) $B^0 \to \chi_{c2} \pi^+ \pi^- K_S^0$,
    (e) $B^0 \to \chi_{c1} \pi^0 \pi^- K^+$ and
    (f) $B^0 \to \chi_{c2} \pi^0 \pi^- K^+$ decay  modes.
    The curves show the signal 
    (red dashed),
    peaking background (magenta dash-dotted) and 
    and the background component (green dotted for 
    combinatorial) as well as the overall fit (blue solid).}
  \label{fig:B_chicJ_Kpipi}
\end{figure*}

$\chicx$ candidates are combined with charged kaon and pion candidate tracks
to reconstruct $B$ meson. To identify the $B$ meson, two kinematical variables
are used:
beam-constrained mass ($M_{\rm bc}$) and energy difference ($\Delta E$).
The $M_{\rm bc}$ is defined as 
$\sqrt{E_{\rm beam}^2 - (\sum_i \vec{p}_{i})^2}$ and the $\Delta E$ 
is defined  as $\sum_i E_i - E_{\rm beam}$, where $E_{\rm beam}$ is the beam 
energy in the CM frame and $p_{i}$ ($E_i$) is the momentum (energy) of 
the $i$-th daughter particle in the CM frame and the
summation is over all 
final states used for reconstruction. We reject candidates having $M_{\rm bc}$ 
less than 5.27~GeV$/c^2$ or $|\Delta E|~>$ 120~MeV.
We extract the signal yield from an unbinned 
extended maximum likelihood (UML) fit to the $\Delta E$ variable.
Fig.~\ref{fig:B_chicJ_Kpi} and~\ref{fig:B_chicJ_Kpipi} shows the fit to
the $\Delta E$ distribution for the decay modes of interest.

To understand the production mechanism of intermediate states,
we look at the background subtracted $M_{\chicx\pi}$,  $M_{K \pi}$, 
$M_{\chicx\pi\pi}$, $M_{K\pi\pi}$, and $M_{\pi\pi}$ distributions for the decay mode of interest.
We perform a UML fit to the $\Delta E$ distribution
and use the $_{S}\mathcal{P}$\textit{lot} 
formalism~\cite{pivk} to project signal events in the distribution of 
interest. 

\begin{figure*}
  \centering
  \includegraphics[trim=1cm 0.5cm 0.25cm 0.25cm,height=90mm,width=80mm]{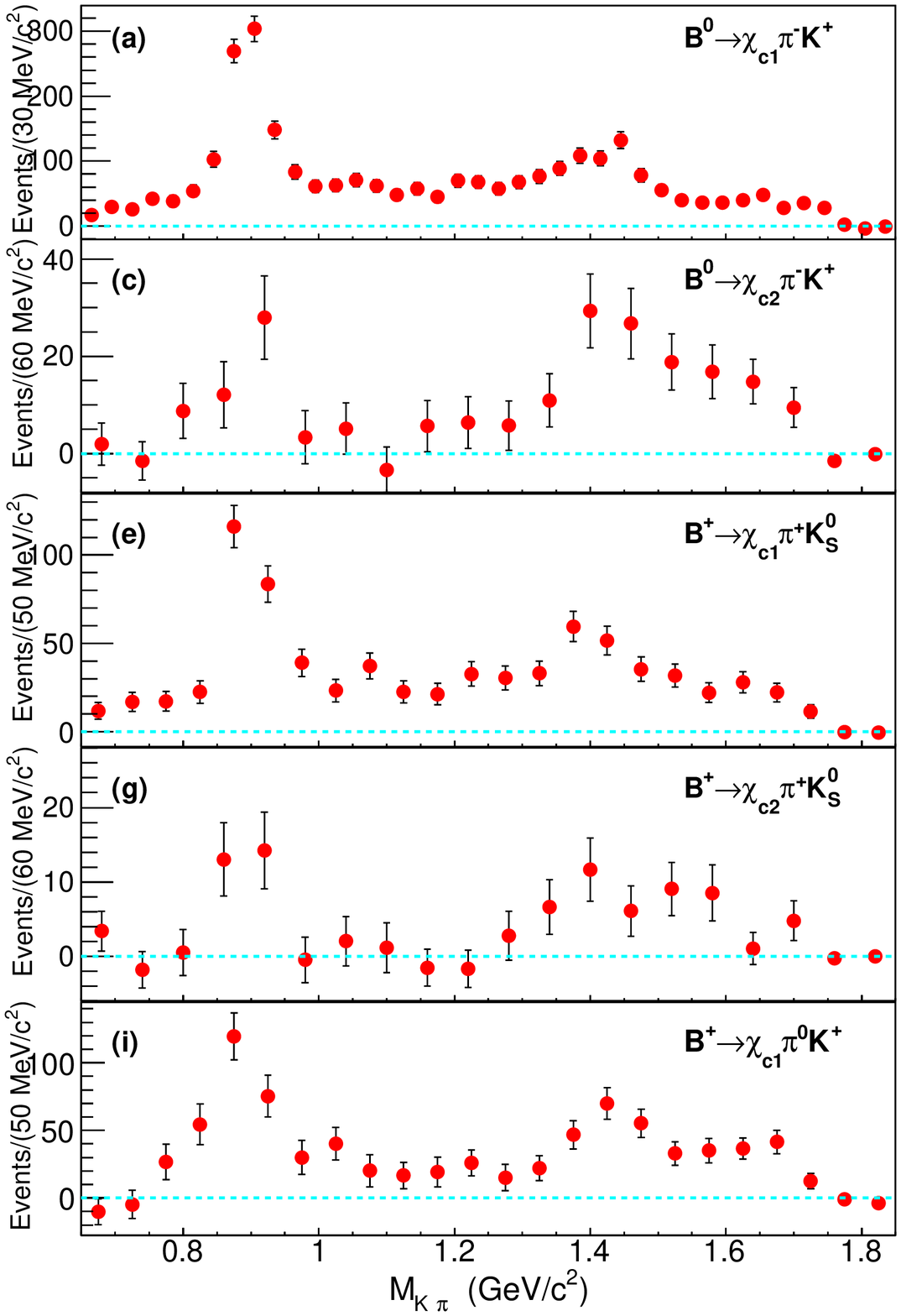}
  \includegraphics[trim=1.25cm 0.5cm 0.25cm 0.25cm,height=90mm,width=80mm]{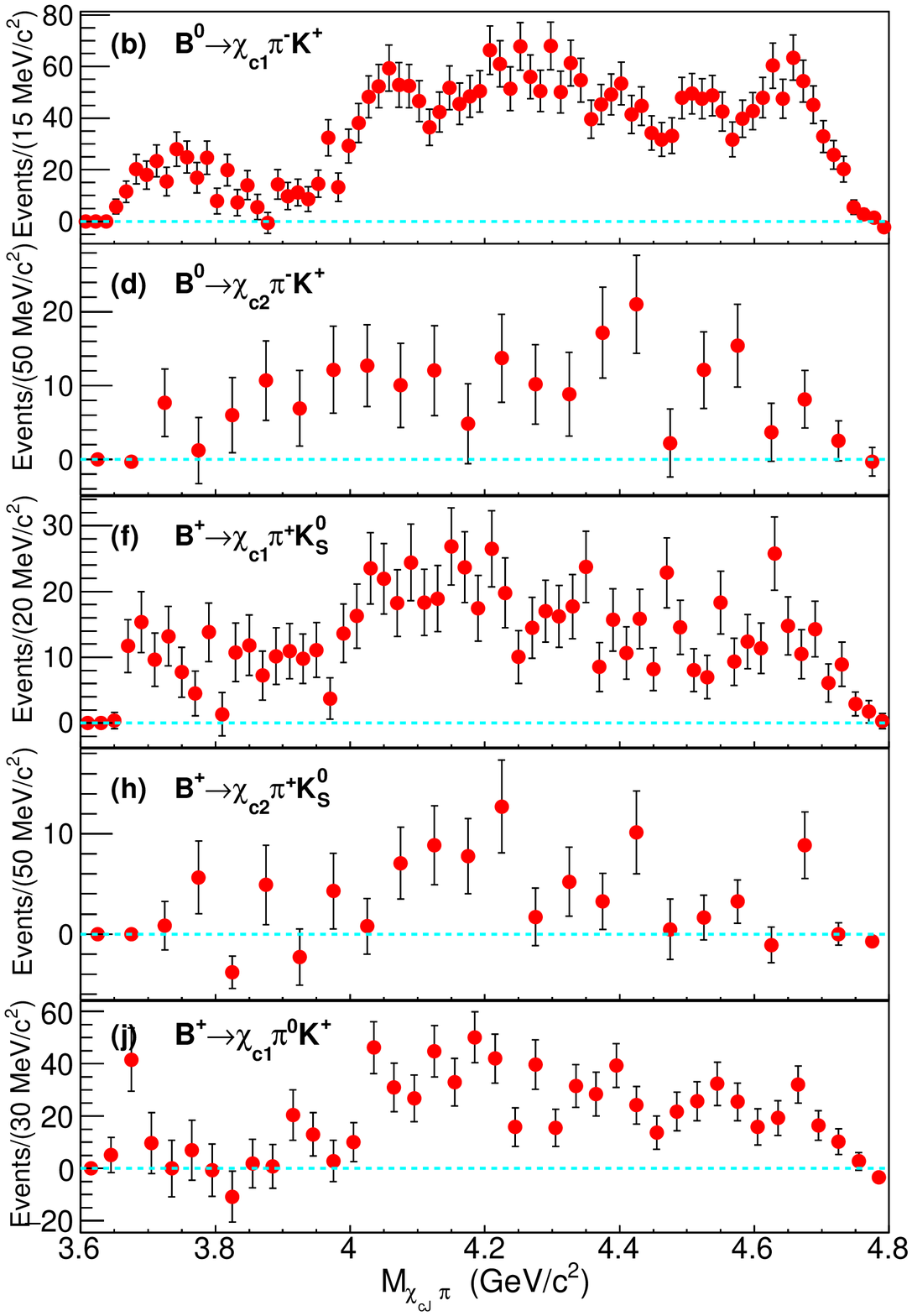}
  
  \caption{Background subtracted 
    $_{S}\mathcal{P}$\textit{lot}
    $M_{K \pi}$ and $M_{\chicx \pi}$ distributions for
    (a and b)  $B^0 \to \chi_{c1} \pi^- K^+$,
    (c and d)  $B^0 \to \chi_{c2} \pi^- K^+$,
    (e and f)  $B^- \to \chi_{c1} \pi^- K_S^0$,
    (g and h)  $B^- \to \chi_{c2} \pi^- K_S^0$ and
    (i and j)  $B^- \to \chi_{c1} \pi^0 K^+$ decay modes.
  }
  \label{fig:Mchipi_all_3body}
\end{figure*}

The $K^{*}(892)$ is found to be a major contribution in the 
$B \to \chi_{c1} \pi K$ decay modes as seen from 
Fig.~\ref{fig:Mchipi_all_3body} (a), (e) and (i), while
in $B \to \chi_{c2} \pi K$ decays  the $K^{*}(892)$ component is less prominent.
Our study suggests that the $B \to \chi_{c2} K^*(892)$ mechanism 
does not 
dominate the $B \to \chi_{c2} \pi K$ decay,  which is in marked contrast to 
the $\chi_{c1}$ case. 
From this study one can say that the production mechanism of the 
$\chi_{c2}$  from $B$ mesons is different in 
three-body decays for the  $B \to  \chicx \pi K $ case.

\begin{figure*}
  \includegraphics[height=65mm,width=80mm]{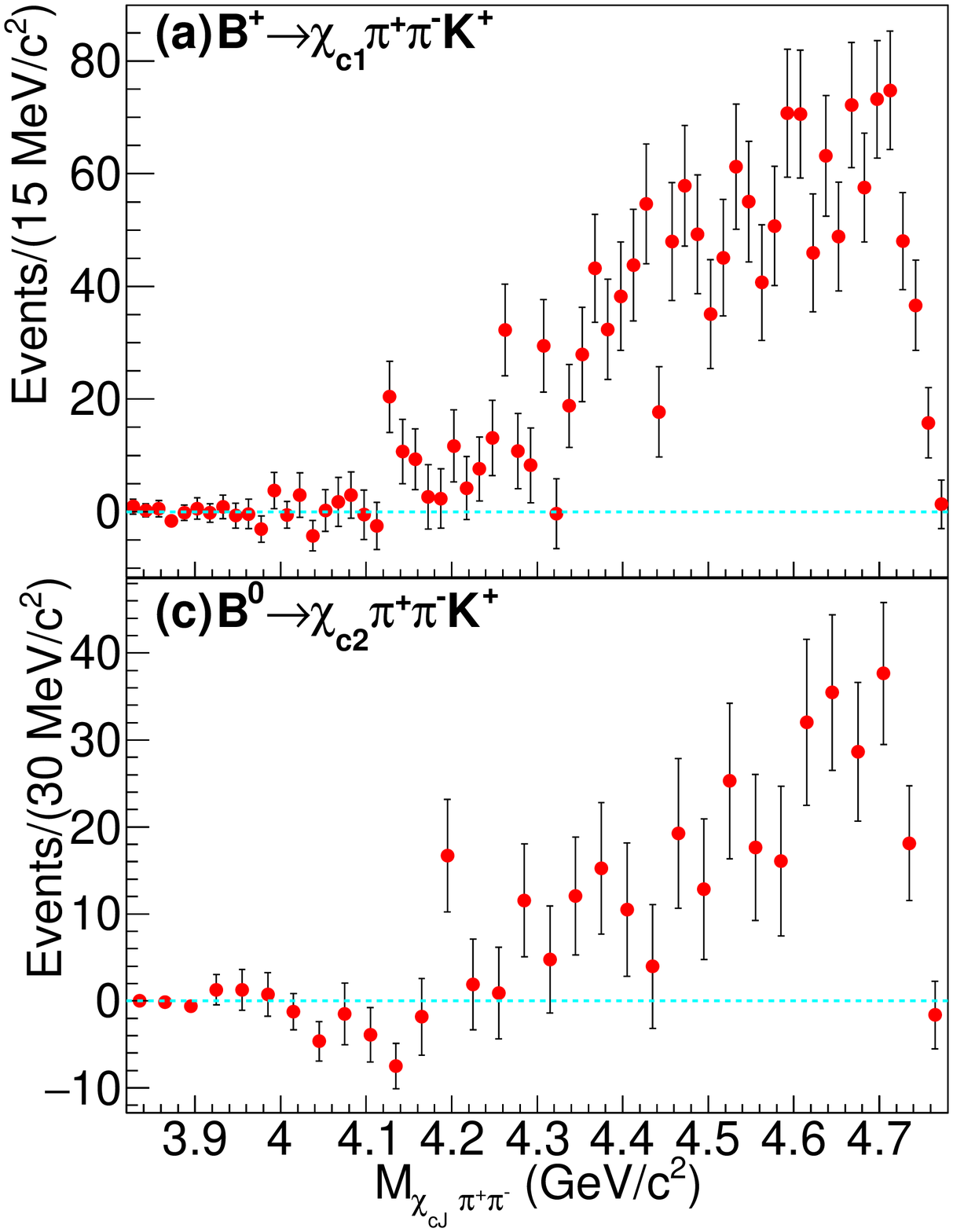}
  \includegraphics[height=65mm,width=80mm]{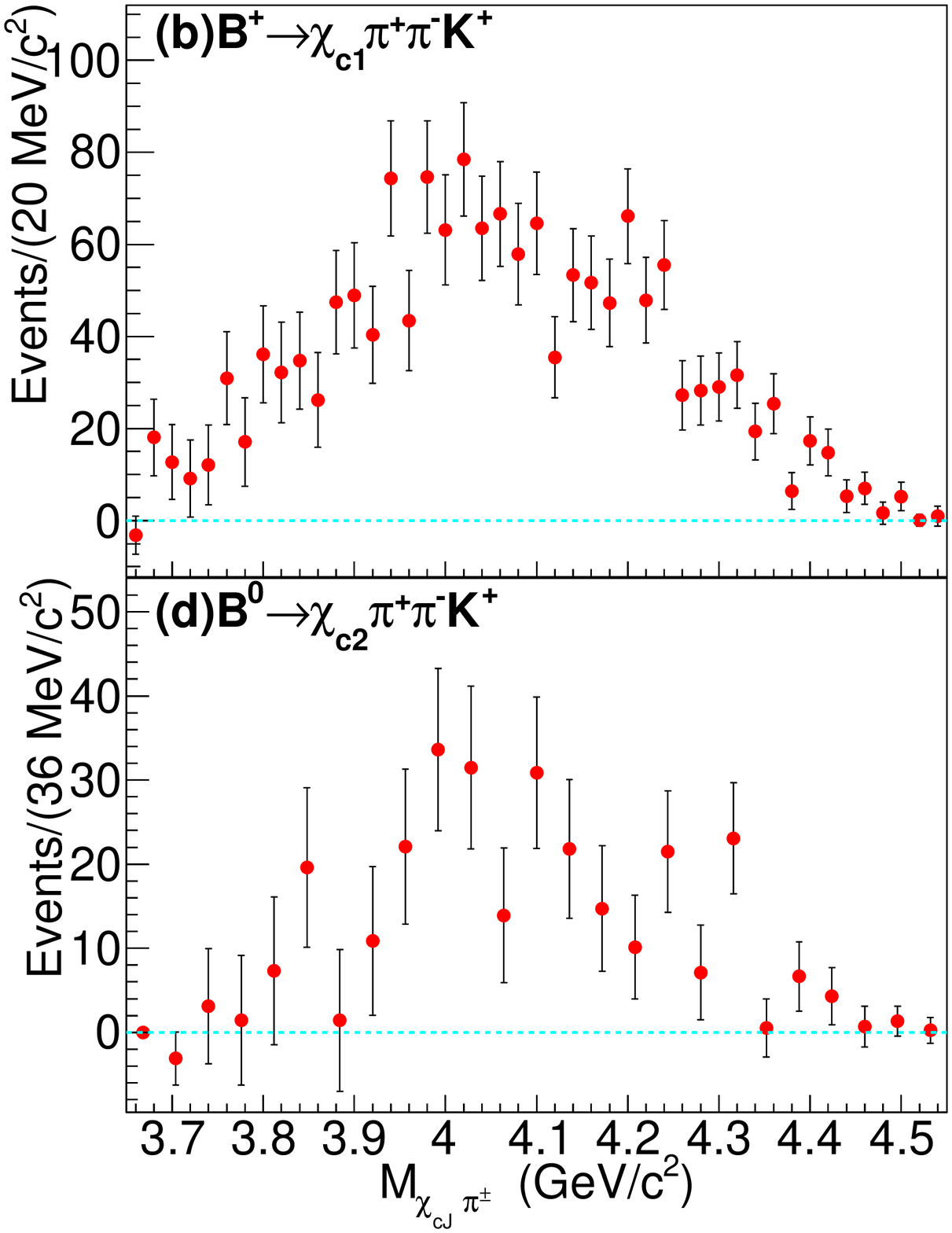}

  \caption{ Background subtracted  $_{S}\mathcal{P}$\textit{lot}
    (a) $M_{\chi_{c1} \pi^+ \pi^-}$,
    (b) $M_{\chi_{c1} \pi^\pm}$,
    (c) $M_{\chi_{c2} \pi^+ \pi^-}$ and   
    (d) $M_{\chi_{c2} \pi^\pm}$ distributions for 
    $B^+ \to \chicx \pi^+ \pi^- K^+$ decay modes.   
    }
  \label{fig:Mchipi_B_chicJ_Kpipi}
\end{figure*}

\begin{figure*}
 \includegraphics[height=65mm,width=50mm]{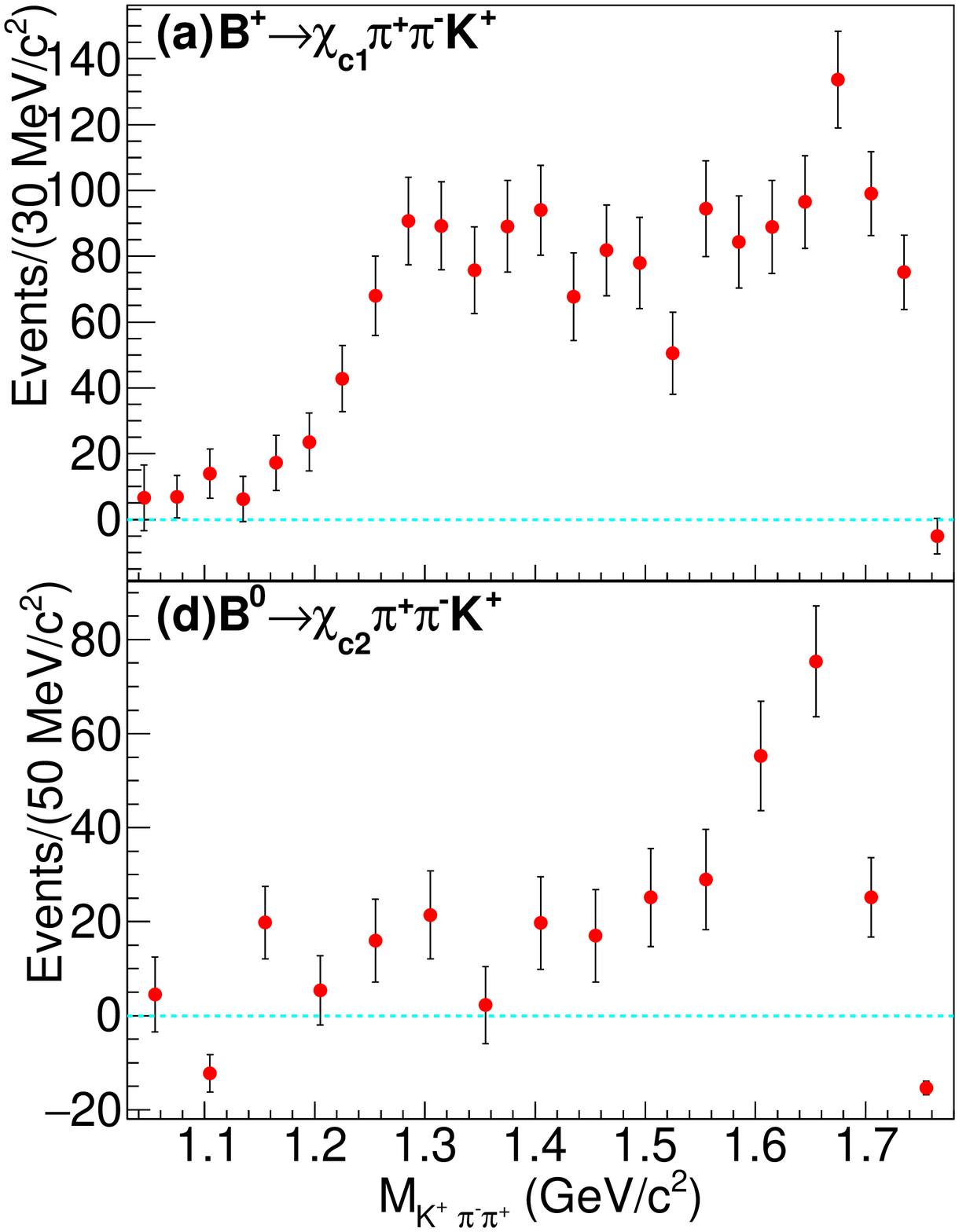}
 \includegraphics[height=65mm,width=50mm]{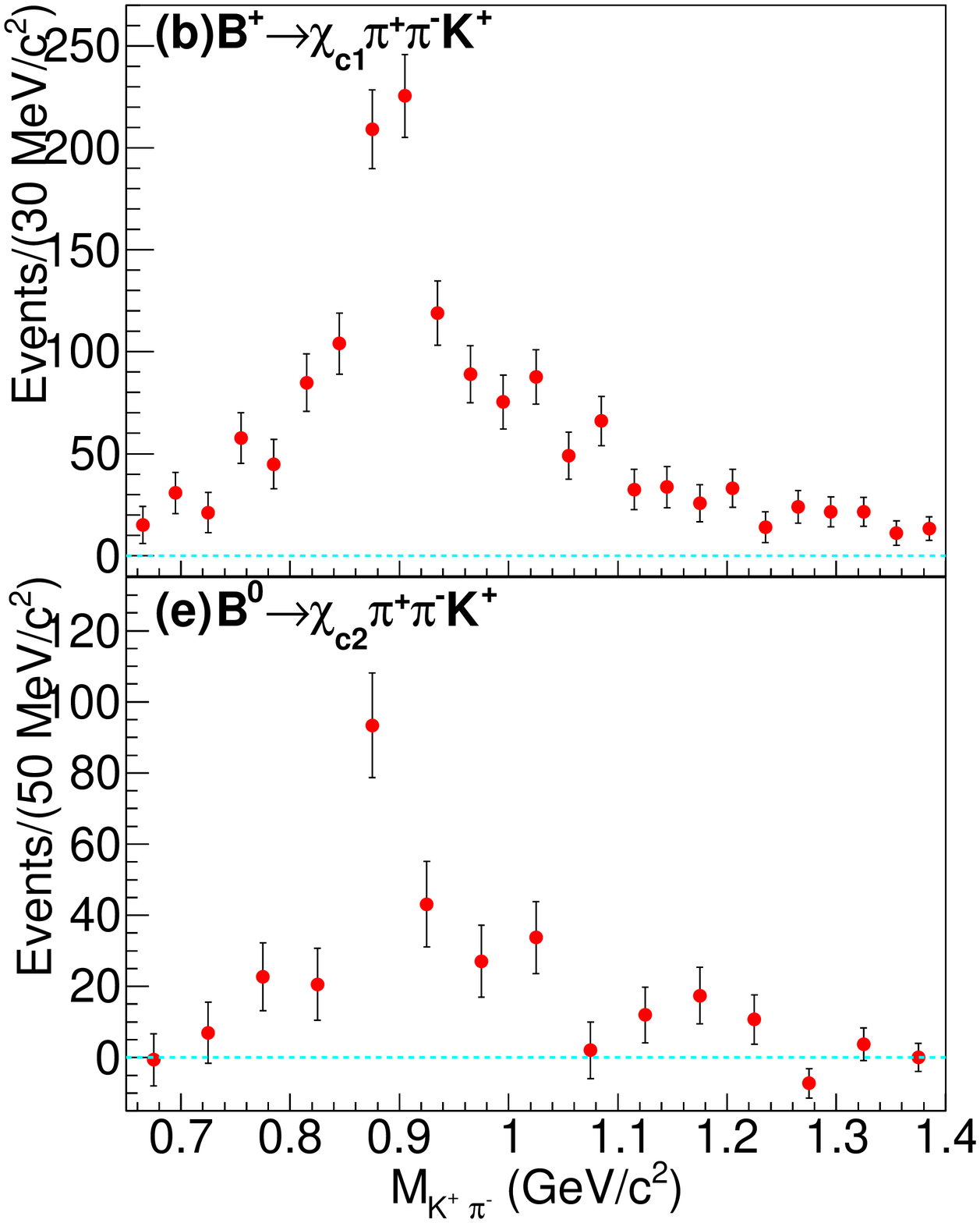}
 \includegraphics[height=65mm,width=50mm]{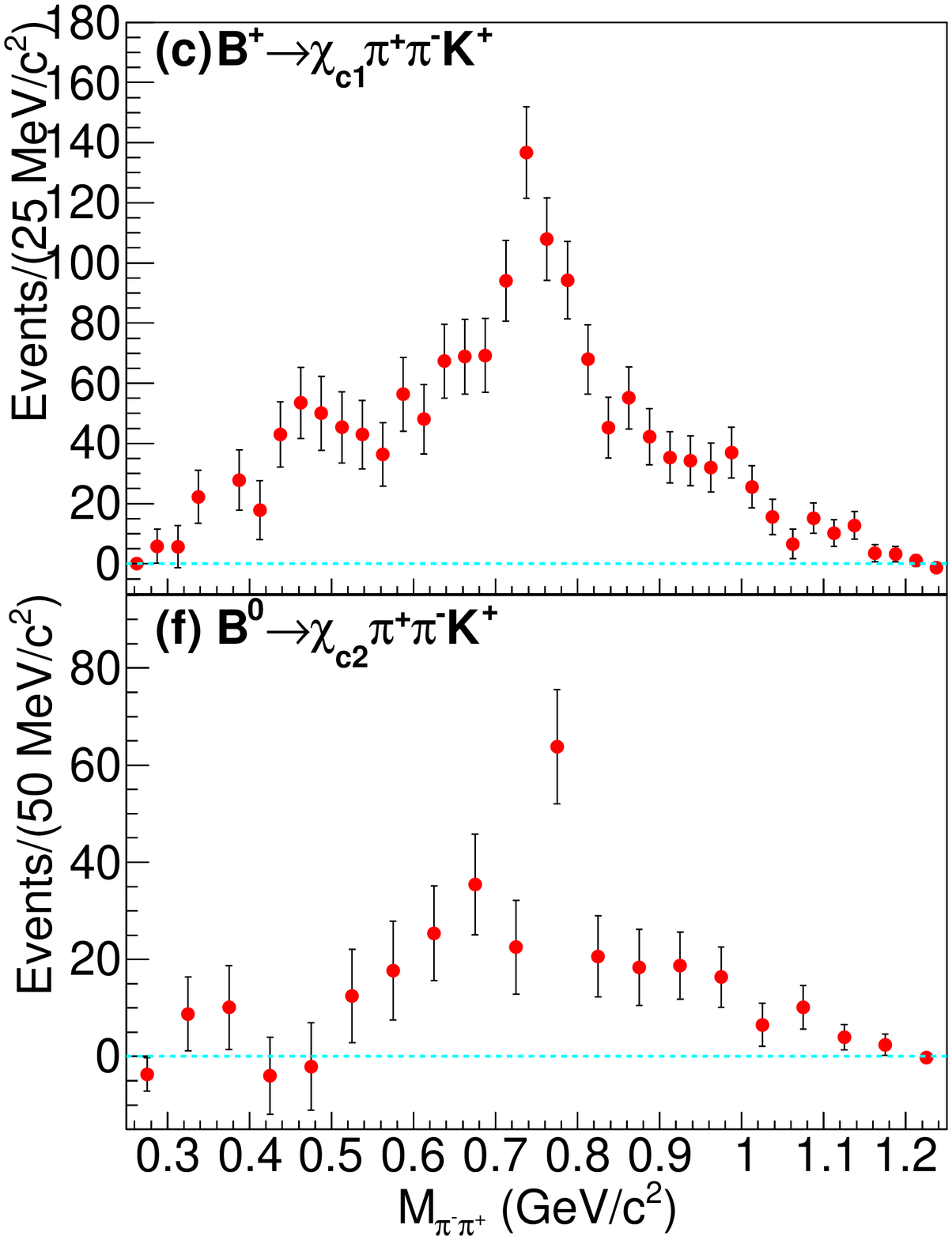}

 \caption{Background subtracted   $_{S}\mathcal{P}$\textit{lot}
    (a and d) $M_{K^+ \pi^+ \pi^-}$,
    (b and e) $M_{K^+ \pi^-}$ and
    (c and f) $M_{\pi^+ \pi^+}$ distributions for 
    $B^+ \to \chi_{c1} \pi^+ \pi^- K^+$ decay (upper) and  
    $B^+ \to \chi_{c2} \pi^+ \pi^- K^+$ decay (lower), respectively.   
    }
  \label{fig:MKpi_B_chicJ_Kpipi}
\end{figure*}

Background subtracted  $_{S}\mathcal{P}$\textit{lot} distribution of
$M_{\chicx \pi \pi}$,   $M_{\chicx \pi^\pm}$,
$M_{K \pi \pi}$, $M_{K^+ \pi^-}$, 
and $M_{\pi^+ \pi^-}$  are shown in
Figs.~\ref{fig:Mchipi_B_chicJ_Kpipi} 
and \ref{fig:MKpi_B_chicJ_Kpipi}
for the $B^+ \to \chicx \pi^+ \pi^- K^+$  decay mode.
No narrow resonance is seen in the $M_{\chicx \pi^+ \pi^-}$   and
$M_{\chicx \pi^\pm}$ distributions with
the current statistics.

\begin{table*}[!htbp]
\caption{ Summary of the results. Signal yield ($Y$) from the fit, 
significance ($\mathcal{S}$) with systematics included,
corrected efficiency ($\epsilon$) and
measured $\mathcal{B}$.
For $\mathcal{B}$, the first (second) error is statistical (systematic). 
 Here, in the neutral $B$ decay case,  the
$K_S^0 \to \pi^+ \pi^-$ branching fraction  is included
in the efficiency ($\epsilon$) but the factor of 2 
(for $K^0 \to K_S^0~{\rm or}~K_L^0$)
is taken into account separately. ${\cal R}_{\cal B}$ is the ratio of $\mathcal{B}(B \to \chi_{c2} X)$ to $\mathcal{B}(B\to \chi_{c1} X)$, where $X$ is the same 
set of particles accompanying the $\chi_{c1}$ ($\chi_{c2}$) in the
final states.
}
\begin{center}

  \begin{tabular}{lcccccc}
\hline \hline
Decay & Yield ($Y$) & $\mathcal{S} (\sigma)$ & $\epsilon$(\%) & ${\cal B}$ $(10^{-4})$ & ${\cal R}_{\cal B}$  \\ \hline

\multicolumn{5}{l}{$B^0 \to\chicx \pi^- K^+$} & $0.14\pm 0.02$
 \\ \hline
 $\chi_{c1}$ & $2774 \pm 66 $ & 66.7 & 17.9  & $4.97\pm0.12\pm0.28$  & \\ 
 $\chi_{c2}$ & $206 \pm 25 $& 8.7 & 16.2   & $0.72\pm0.09\pm0.05$ & \\ \hline
 
 \multicolumn{5}{l}{$B^+ \to \chicx \pi^+ K^0$}  & $0.20\pm 0.04$  
 \\ \hline
 $\chi_{c1}$ & $770 \pm 35$  &  33.7 &  8.6 &  $5.75\pm 0.26 \pm 0.32$ & \\ 
 $\chi_{c2}$ &  $76.4 \pm 14.7 $ &  4.6 & 7.5 & $1.16\pm 0.22\pm0.12$&  \\ \hline

\multicolumn{5}{l}{$B^+ \to\chicx \pi^0 K^+$} &   $<0.21$
\\ \hline
$\chi_{c1}$ & $803 \pm 70$ & 15.6 & 7.8 & $3.29\pm0.29\pm0.19$ & \\
$\chi_{c2}$ & $17.5\pm 28.4 $ & 0.4 & 7.0 & $<0.62$ & \\ \hline


\multicolumn{5}{l}{$B^+ \to \chicx \pi^+ \pi^- K^+$}  & $0.36\pm0.05$   
\\ \hline
$\chi_{c1}$ & $1502 \pm 70$ & 19.2 & 12.8 & $3.74 \pm 0.18 \pm0.24 $ & \\
$\chi_{c2}$ & $269 \pm 34$ & 8.4 & 11.4 & $1.34\pm0.17\pm0.09$ & \\ 
\hline

\multicolumn{5}{l}{$B^0 \to \chicx \pi^+ \pi^- K^0$}  & $<0.61$
\\ \hline
$\chi_{c1}$ & $268 \pm 30 $ &  7.1 & 5.4 &  $3.16\pm0.35\pm0.32$ & \\
$\chi_{c2}$ &  $37.8 \pm 14.2 $ & 1.8 & 4.8 &  $<1.70$  & \\ 
\hline

\multicolumn{5}{l}{$B^0 \to \chicx \pi^0 \pi^- K^+$}  & $<0.25$ 
\\ \hline
$\chi_{c1}$ & $545 \pm  81$ &6.5 & 5.0  & $3.52 \pm 0.52 \pm0.24$ & \\
$\chi_{c2}$ & $-76.7 \pm 42.0 $ & - & 4.3 & $<0.74$  &\\  \hline
\hline

\end{tabular}
\label{tab:final_results}
\end{center}
\end{table*}

In our search for the  $X(3872)$ and/or $\chi_{c1}(2P)$ decaying into 
$\chi_{c1} \pi^+ \pi^-$, we didn't find any signal and provided upper limit
(@ 90\% C.L.) as:
\begin{itemize}
\item{ $\mathcal{B}(B^{\pm}\to X(3872)K^{\pm}) \times 
\mathcal{B}(X(3872)\to\chi_{c1}\pi^+\pi^-) < 1.4 \times 10^{-6}$}
\item{$\mathcal{B}(B^+\to\chi_{c1}(2P)K^+)\times \mathcal{B}(\chi_{c1}(2P)\to\chi_{c1}(1P)\pi^+\pi^-) < 1.1 \times 10^{-5}$}
\end{itemize}

\section{Summary}
Belle presented the preliminary results at this conference.
We measured the direct 
branching fractions 
$\mathcal{B}(B\to \chi_{c1} X)$ and
$\mathcal{B}(B\to \chi_{c2}X)$  to be
$(3.03 \pm 0.05\pm 0.25)\times 10^{-3}$ and
$(0.70\pm 0.06\pm0.10)\times 10^{-3}$, respectively.
We observe the $B^0 \to \chi_{c2} \pi^- K^+$ decay mode 
for the first  time, with $206\pm25$ signal events having a 8.7$\sigma$ 
significance,  along with evidence for the 
$B^+ \to \chi_{c2} \pi^+ K_S^0$ decay  mode, with  76$\pm$15 signal events
and a significance of $4.6\sigma$.
In four-body decays, we observe the 
$B^+ \to \chi_{c1} \pi^+ \pi^- K^+$, $B^+ \to \chi_{c1} \pi^+ \pi^- K^+$,
$B^0 \to \chi_{c1}  \pi^+ \pi^- K_S^0$,  and
$B^0 \to \chi_{c1}  \pi^0 \pi^- K^+$ decay modes for the first time 
and report on measurements 
of their branching fractions.  We find that  
$\chi_{c2}$ production in comparison with $\chi_{c1}$ increases 
with a higher number of multi-body $B$ decays.
We observe that the $\chi_{c2}$ is more often accompanied by 
higher $K^*$ resonances as opposed the $\chi_{c1}$ which is dominantly 
produced with lower $K^*$ resonance.
Inclusive and exclusive study of $B$ decays having $\chi_{c2}$ 
in the final state suggests suppression of two-body $B$ decay 
due to suppression of a tensor,  while multi-body $B$ decays 
into $\chi_{c2}$ are allowed.
In our search for $X(3872) \to \chi_{c1} \pi^+ \pi^-$ and
$\chi_{c1}(2P)$, we determine an U.L. on 
the product of branching fractions. The 
negative result for our searches is compatible with the interpretation of
$X(3872)$ as an admixture state of a 
$D^0 \bar{D}^{*0}$ molecule and a $\chi_{c1}(2P)$ charmonium state.

\Acknowledgements
We like to thank KEKB and all members of Belle collaboration along with their 
supporting funding agencies.
This work is supported by a Grant-in-Aids from MEXT for Scientific Research on
Innovative Areas (``Elucidation of New Hadrons with a Variety of Flavors'')
and the U.S. Department of Energy and the National Science Foundation.


\end{document}

%% file: econfmacros.tex
\def\beq{\begin{equation}}
\def\eeq#1{\label{#1}\end{equation}}
\def\eeqn{\end{equation}}

\def\beqa{\begin{eqnarray}}
\def\eeqa#1{\label{#1}\end{eqnarray}}
\def\eeqan{\end{eqnarray}}

\let\bar=\overbar

\def\Dslash{\not{\hbox{\kern-4pt $D$}}}
\def\dslash{\not{\hbox{\kern-2pt $\del$}}}

\def\msb{{\bar{\ssstyle M \kern -1pt S}}}




%% file: chic.bbl
\begin{thebibliography}{99}
\bibitem{CX}
Hereinafter, $\chicx$ refers to  either $\chi_{c1}$ or $\chi_{c2}$, 
depending  on  which is reconstructed.
\bibitem{mixchg}
Charge-conjugate and neutral modes are included throughout the paper 
unless stated otherwise.
\end{thebibliography}
